\documentclass[12pt]{article}
\usepackage[applemac]{inputenc}
\usepackage{amsmath,amssymb}
\usepackage{paralist}
\usepackage{slashed}
\usepackage{graphicx}
\usepackage{subfigure}
\usepackage{color}
\usepackage{fullpage}
\usepackage{multirow}
\usepackage{hyperref}
\usepackage{amscd}
\usepackage{xcolor}
\usepackage{braket}
\usepackage{amsfonts}

\catcode`\@=11 
\def\numberbysection{\@addtoreset{equation}{section} 
        \def\theequation{\thesection.\arabic{equation}}} 

\def\be{\begin{equation}} 
\def\ee{\end{equation}} 
\def\ba{\begin{eqnarray}} 
\def\ea{\end{eqnarray}} 
\def\bali{\begin{align}}

\def\nl{\nonumber \\}

\def\de{\partial}

\def\dag{\dagger} 
 
\def\a{\alpha} 
\def\b{\beta} 
\def\g{\gamma} 
\def\G{\Gamma} 
\def\D{\Delta} 
\def\d{\delta} 
\def\e{\epsilon} 
\def\eps{\varepsilon} 
 
\def\h{\eta} 
 
\def\l{\lambda} 
 
\def\m{\mu} 
\def\n{\nu}

\def\r{\rho} 
\def\s{\sigma}

\def\f{\phi}

\def\c{\chi} 
\def\w{\omega}

\def\th{\theta}

\def\u1{\widehat{U(1)}}
\def\su2{\widehat{SU(2)}_1}

\begin{document} 
 
\begin{titlepage} 
\begin{center} 
 
\vskip .6 in 
{\LARGE Multipole Expansion }
\medskip

{\LARGE in the Quantum Hall Effect} 
 
\vskip 0.2in 
Andrea CAPPELLI ${}^{(a)}$\\ 
 
Enrico RANDELLINI ${}^{(a,b)}$\\ 
 \medskip

{\em ${}^{(a)}$ INFN, Sezione di Firenze}\\ 
{\em ${}^{(b)}$ Dipartimento di Fisica, Universit\`a di Firenze\\ 
Via G. Sansone 1, 50019 Sesto Fiorentino - Firenze, Italy} 
\end{center} 
\vskip .2 in 
\begin{abstract} 
The effective action for low-energy excitations of Laughlin's states is 
obtained by systematic expansion in inverse powers of the magnetic 
field. It is based on the W-infinity symmetry of quantum incompressible 
fluids and the associated  higher-spin fields.
Besides reproducing the Wen and Wen-Zee actions and the Hall viscosity, 
this approach further indicates that the low-energy excitations are extended 
objects with dipolar and multipolar moments.
\end{abstract} 
 
%\pacs{73.43.Cd, 11.25.Hf, 73.23.Hk, 73.43.Jn} 
\vfill 
 
\end{titlepage} 
\pagenumbering{arabic} 
\numberbysection
 
%-1--------------------------------------------- 
 
\section{Introduction} 

Many authors have recently reconsidered the Laughlin theory
of the quantum Hall incompressible fluid \cite{GP} aiming at
understanding it more deeply and obtaining further universal
properties, often related to geometry.  The system has been considered
on spatial metric backgrounds for studying the heat transport
\cite{heat} and the response of the fluid to strain.  In particular, the
Hall viscosity has been identified as a new universal quantity
describing the non-dissipative transport \cite{Avron} \cite{Read}.

Some authors have also been developing physical models of the Hall
fluid that go beyond the established picture of Jain's composite
fermion.  Haldane and collaborators have considered the response of
the Laughlin state to spatial inhomogeneities (such as lattice effects
and impurities) and have introduced an internal metric degree of
freedom, that suggests the existence of dipolar effects
\cite{Haldane}.  Wiegmann and collaborators have developed an
hydrodynamic approach describing the motion of a fluid of electrons as
well as that of vortex centers \cite{Wiegmann-vortex}.

In the study of the quantum Hall system, the low-energy effective action has
been a very useful tool to describe and parameterize physical effects,
and to discuss the universal features. Besides the well-known Chern-Simons
term leading to the Hall current, the coupling to gravity was
introduced by Fr\"ohlich and collaborators \cite{Fro} and by Wen and
Zee \cite{Wen-Zee}.  The resulting Wen-Zee action describes the Hall
viscosity and other effects in term of the parameter $\bar s$,
corresponding to an intrinsic angular momentum of the low-energy
excitations. This quantity, independent of the relativistic spin,
suggests a spatially extended structure of excitations. The
predictions of the Wen-Zee action have been checked by the microscopic
theory of electrons in Landau levels (in the case of integer Hall
effect \cite{Abanov1}) and corrections and improvements have been
obtained \cite{Fradkin-geo}\cite{Abanov-framing}.  Further
features have been derived under the assumption of local Galilean
invariance of the effective theory
\cite{Son1}\cite{Son2}\cite{Abanov2}\cite{Abanov-boundary}.

In this paper, we rederive the Wen-Zee action by using a different
approach that employs the symmetry of Laughlin incompressible fluids
under quantum area-preserving diffeomorphism ($W_{\infty}$ symmetry)
\cite{W1}.  The consequences of this symmetry on the dynamics of edge
excitations have been extensively analyzed \cite{CDTZ}; in particular,
the corresponding conformal field theories have been obtained and
shown to characterize the Jain hierarchy of Hall states \cite{W2}.
Regarding bulk excitations, the $W_{\infty}$ symmetry and the
associated effective theory have not been developed, were it not for
the original studies by Sakita and collaborators \cite{Sakita} and the
classic paper \cite{GMP}.

Here, we study the bulk excitations generated by $W_{\infty}$
transformations in the lowest Landau level.  We disentangle their
inherent non-locality by using a power expansion in $(\hbar/B_0)^n$,
where $B_0$ is the external magnetic field. Each term of this
expansion defines an independent hydrodynamic field of spin
$\s=1,2,\cdots$, that can be related to a multipole amplitude of the
extended structure of excitations. The first term is just the Wen
hydrodynamic gauge field, leading to the the Chern-Simons action
\cite{Wenbook}.  The next-to-leading term involves a traceless
symmetric two-tensor field, that is a kind of dipole moment.  Its
independent coupling to the metric background gives rise to the
Wen-Zee action and other effects found in the literature.  The
third-order term is also briefly analyzed. The structure of this
expansion matches the non-relativistic limit of the theory of
higher-spin fields in $(2+1)$ dimensions and the associated
Chern-Simons actions developed in the Refs.\cite{HS}.

Our approach allows to discuss the universality of
quantities related to transport and geometric responses.
We argue that the general expression of the effective action contains 
a series of universal coefficients, the first of which is the Hall 
conductivity and the second is the Hall viscosity.
We also identify other terms that are not universal because
they correspond to local deformations of the bulk effective action.
In principle, all the universal quantities can be observed once 
we probe the system with appropriate background fields, but so 
far our analysis is complete to second order in $\hbar/B_0$ only.

We believe that the multipole expansion offers
the possibility of matching with the physical models of dipoles and
vortices by Haldane and Wiegmann mentioned before \cite{Haldane}
\cite{Wiegmann-vortex}. Moreover, in our approach, the intrinsic
angular momentum $\bar s$ receives a natural interpretation.

The paper is organized as follows. In section two, we recall the
original derivation of the Wen-Zee action \cite{Wen-Zee}.  We spell
out the major physical quantities obtained from this action, using
formulas for curved space that are summarized in Appendix A.  In
section three, we present the basic features of the $W_\infty$ symmetry
on the edge and in the bulk; we set up the $\hbar/B_0$ expansion and
introduce the associated higher-spin hydrodynamic fields. The coupling
to the electromagnetic and gravity backgrounds of the first two fields
is shown to yield the Wen-Zee action. Next, the issue of universality
of the effective action is discussed. Then, the third-order field is
introduced and its contribution to the effective action is found.  In
section four, the physical picture of dipoles is described
heuristically. In the Conclusions, some developments of this
approach are briefly mentioned.

%-2--------------------------------------------------
\section{The Wen-Zee effective action}

We consider the Laughlin state with filling fraction $\nu=1/p$ and
density $\r_0=\nu B_0/2\pi$ (setting $\hbar=c=e=1$ for convenience).
The matter fluctuations are described by the conserved current $j^\mu$,
with vanishing ground state value, that is expressed in terms of the 
hydrodynamic U(1) gauge field $a_\mu$ ($\mu=0,1,2$),
\be
j^\mu=\eps^{\mu\nu\r}\,\de_\nu a_\r\ ,
\label{current1}
\ee
where $\eps^{\mu\nu\r}$ is the antisymmetric symbol, $\eps^{012}=1$.
The leading low-energy  dynamics for this gauge field compatible
with the symmetries of the problem is given the Chern-Simons term, leading
to the effective action \cite{Wenbook}:
\be
S[a,A]= \int \r_0 A_0 +\int -\frac{1}{2\g} ada +j^\mu A_\mu\ .
\label{Seff1}
\ee
In this equation, we introduced the coupling to the external 
electromagnetic field $A_\mu$, we included the static contribution and
used the short-hand notation of differential forms, $a=a_\mu dx^\mu$.

Integration of the hydrodynamic field leads to the induced action
$S_{\rm ind}[A]\equiv S[A]$, that expresses the response of the system to
the electromagnetic background,
\be
S[A]=\frac{\nu}{4\pi}\int AdA, \qquad \qquad \nu=\frac{1}{p}.
\label{CS}
\ee
Its variations yield the density and Hall current, 
\ba
\r&=&\frac{\d S}{\d A_0}=\frac{\nu}{2\pi}{\cal B}=
\frac{\nu}{2\pi}\left(B_0+\d B(x)\right),
\label{rho1}
\\
J^i&=&\frac{\d S}{\d A_i}=\frac{\nu}{2\pi}\eps^{ij} {\cal E}^j,
\label{curr1}
\ea
where ${\cal B}$ and ${\cal E}^i$ are the magnetic and electric
fields, respectively. The Chern-Simons coupling constant in (\ref{Seff1})
has been identified as $\g=\nu/2\pi$. 
As is well-known \cite{Wenbook}, the Chern-Simons
theory (\ref{Seff1}) describes local excitations of the
$a_\mu$ field that possess fractional statistics with parameter
$\th=\pi/p$. Moreover, the action is not gauge invariant
and a boundary term should be added; this is the $(1+1)$ dimensional
action of the chiral boson theory (chiral Luttinger liquid) that
realizes the conformal field theory of edge excitations.

The Wen-Zee action is obtained by coupling the hydrodynamic
field to a spatial time-dependent gravity background, as follows 
\cite{Fro}\cite{Wen-Zee}. The metric takes the form:
\be
g_{ij}(t,x^k)=e^a_i e^b_j \,\d_{ab}, \qquad i,j,k,a,b=1,2,
\label{metric}
\ee
also written in terms of the zweibein $e^a_i$. 
Note that we do not introduce
time and mixed components of the metric, $g_{00}=g_{0i}=0$, such that
the resulting theory will only be covariant under time-independent
reparameterizations. Actually, we shall find
non-covariant time-dependent effects that are physically relevant.
We also assume that the gravity background has vanishing torsion, 
such that the metric and zweibein
descriptions are equivalent; in particular, the spin connection 
$\w_\mu^{ab}$ and Levi-Civita connection $\G^i_{jk}$ 
describe equivalent physical effects. In Appendix A,
we summarize some useful formulas of covariant calculus.

The comoving coordinates are invariant under local
O(2) rotations and the corresponding spin connection is an Abelian
gauge field, $\w_\mu=\w_\mu^{ab}(e)\eps_{ab}/2$. 
The standard coupling of the spin connection 
to the spin current of the relativistic fermion in $(2+1)$ dimension
has the following non-relativistic limit (A,B=0,1,2):
\be
\w^{AB}\, S^\mu_{AB} =\w^{AB}\, \bar\psi\g^\mu
\frac{1}{4}\left[\g_A,\g_B \right]\psi \ \longrightarrow\ 
\frac{1}{2}\w^{12}_\mu\ \bar\psi\g^\mu\s^3\psi
\ \sim\ \frac{1}{2}\w^{12}_\mu\ \bar\psi\g^\mu\psi,
\label{relativistic-sc}
\ee
namely, it reduces to the charge interaction.
This result suggests to introduce the following coupling to gravity
in the effective action (\ref{Seff1}),
\be
j^\mu A_\mu \ \longrightarrow j^\mu \left( A_\mu+ \bar s\, \w_\mu\right),
\label{grav-coupl}
\ee
where $\bar s$ is a free parameter measuring the intrinsic angular
momentum of low-energy excitations. 
The resulting induced action, generalizing (\ref{CS}), reads
\cite{Wen-Zee}:
\be
S[A,g]=\frac{\nu}{4\pi}\int AdA + 2\bar{s}\, Ad\w +\bar{s}^2 \, \w d\w \,.
\label{CS-WZ}
\ee
In this expression, the second term is usually referred as the Wen-Zee action, 
$S_{WZ}[A,g]$, while the third part $O(\bar{s}^2)$
is called `gravitational Wen-Zee term', $S_{GRWZ}[g]$.

The effective action (\ref{CS-WZ}) is the main quantity of our analysis
in this work. The coupling to the spin connection (\ref{grav-coupl}) has been
confirmed by the study of world lines of excitations in $(2+1)$
dimensions \cite{Fradkin-geo}. 
Moreover, the correctness of the action (\ref{CS-WZ})
has been verified by integrating the microscopic theory of electrons 
in Landau levels, for integer $\nu$ \cite{Abanov1}.
These works have noted that there is a contribution
from the measure of integration of the path integral over
$a_\mu$; this is the framing (gravitational) anomaly of the
Chern-Simons theory \cite{Abanov-framing}, and leads to an 
additional Wess-Zumino-Witten
term in the effective action. This yields a redefinition of the
coefficient of $S_{GRWZ}$, $\bar{s}^2\ \to\ \bar{s}^2 -c/12$,
where $c$ is the central charge of the conformal theory on the
boundary (i.e. $c=1$ for Laughlin states).
Note that the bar in $\bar{s}$, indicating the average over the contribution
of several Landau levels, is not actually relevant for Laughlin states, such
that $\bar{s}^2=\bar{s^2}$ in the following discussion.

In a actual system, the effective action (\ref{CS-WZ}) is accompanied by
other non-geometrical terms that are local and gauge invariant and
depends on the details of the microscopic Hamiltonian 
\cite{Son1}\cite{Abanov1}. These
non-universal parts will not be considered here, while
the issue of universality will be discussed later.

In the following, we review the physical consequences that can be
obtained from the first two terms in the action (\ref{CS-WZ})
and postpone the analysis of the gravitational part $S_{GRWZ}[g]$ to 
Section 3.5. The Wen-Zee action involves three terms,
\be
\label{WZ3}
S_{WZ}=\frac{\n\bar{s}}{2\pi} \int A d \w=
\frac{\n\bar{s}}{2\pi}\int d^3x \left(\frac{\sqrt{g}}{2} A_0 \mathcal{R} + 
 \e^{ij} \dot{A_i} \w_j + \sqrt{g}\,{\cal B}\, \w_0 \right)\ ,
\ee
where we introduced the scalar curvature of the spatial metric and
the total magnetic field through the expressions (cf. Appendix A):
\be
{\cal R} =\frac{2}{\sqrt{g}}\eps^{ij}\de_i\w_j\ ,
\qquad
{\cal B} =\frac{1}{\sqrt{g}}\eps^{ij}\de_iA_j\ .
\label{ScalarR}
\ee

From the variation of the effective action with respect to $A_0$ we
obtain a contribution to the density that is proportional to the
scalar curvature; this is relevant when the system 
is put on a curved space, such as e.g. the sphere.
Integrating the density over the surface, we find that the total
number of electrons is:
\be
\label{shift}
N=\int d^2x \sqrt{g} \r= \frac{\n}{2\pi}\int d^2x\ \sqrt{g}
\left( {\cal B} + \frac{\bar s}{2} \mathcal{R} \right)= \n N_{\f} +  
\n\bar{s} \c =  \n N_{\f} +  \n\mathcal{S},
\ee
where $N_{\f}$ is the  number of magnetic 
fluxes going through the surface and $\c$ 
is its Euler characteristic. This relation shows that on a curved space the 
number of electrons $N$ and the number of flux quanta $N_{\f}$ 
are not simply related by $N=\n N_{\f}$.
Rather there is a sub-leading $O(1)$ correction, called 
the shift $\mathcal{S}=\bar{s} \c$. For the sphere, this 
is $\mathcal{S}=2\bar{s}$; upon comparing with the actual expression 
of the Laughlin wave function in this geometry, one obtains
the value of the intrinsic angular momentum $\bar{s}=1/2\n=p/2$ 
\cite{Wen-Zee}. 

The shift is another universal quantum number characterizing 
Hall states, besides Wen's topological order \cite{Wenbook},
that depends on the topology of space.
One simple way to compute $\bar{s}$ is to consider
the  total angular momentum $M$ of the ground state wavefunction
for $N$ electrons and use the following formula:
\be
M=\frac{N}{2}N_\f = \frac{N^2}{2\nu}- N \bar{s}.
\label{ang-mom}
\ee
The sub-leading $O(N)$ term in this expression gives 
the intrinsic angular momentum $\bar s$ of excitations.
For the $n$-th filled Landau level one finds $\bar s=(2n-1)/2$; 
in the lowest level, for wavefunctions given by conformal field theory,
Read has obtained the general formula, 
$\bar s= 1/2\nu+h_\psi$, where $h_\psi$ is the scale dimension
of the conformal field $\psi$ representing the neutral part of the
electron excitation \cite{Read}.

The induced Hall current obtained from the variation of the action (\ref{WZ3}) 
with respect to $A_i$ reads:
\be
J^i=\frac{1}{\sqrt{g}} \frac{\d S[A,g]}{\d A_i}=\frac{1}{\sqrt{g}} 
\frac{\n}{2\pi}\e^{ij}\left( {\cal E}^j + \bar{s}\, {\cal E}_{(g)}^j \right),
\qquad {\cal E}_{(g)}^i=\de_i\w_0-\de_0\w^i,
\label{grav-E}
\ee
and shows a correction given by the `gravi-electric'
field ${\cal E}^i_{(g)}$.

The most important result of the Wen-Zee action is given by the purely
gravitational response encoded in the third term of (\ref{WZ3}).  For
small fluctuations around flat space, $g_{ij}=\d_{ij} + \d g_{ij}$,
the metric represents the so called strain tensor of elasticity
theory, $\d g_{ij}=\de_i u_j + \de_j u_i$, where $u_i (x)$ is the local
deformation \cite{Landau}.  In order to find the response of the fluid
to strain, we should compute the induced stress tensor. To this
effect, we expand the Wen-Zee action for weak gravity and rewrite it
explicitly in terms of the metric.

The relation between the metric and the zweibeins 
(\ref{metric}) can be approximated as follows. We choose a gauge for the local
O(2) symmetry such that the zweibeins form a symmetric matrix.
Then, to leading order in the fluctuations we can write,
$\d g_{ij}=\d e^a_j \d_{ai} +\d e^a_i \d_{aj} =2 \d e_{ij}$, and
express the zweibeins in terms of the metric. 
The  spin connection components are then found to be (see Appendix A):
\be
\w_0= -\frac{1}{8}\eps^{ik}\, \d g_{ij}\, \d \dot{g}_{kj},
\qquad\quad
\w_j=\frac{1}{2}\eps^{ki}\, \de_i \d g_{kj},
\label{w0-wi-linear}
\ee
where the dot indicates the time derivative. Note
that $\w_0$ and $\w_i$ are quadratic and linear in the metric 
fluctuations, respectively.
Moreover, to linear order the spatial zweibein is
proportional to the affine connection,
\be
\w_i= \frac{1}{2}\G_i\equiv\frac{1}{2}\eps^{jk}\, \G_{j,ik},
\label{w-gamma}
\ee
and the curvature reads:
\be
\mathcal{R}=\e^{ij} \, \de_i \G_j=
\left(\de_i\de_j - \d_{ij}\de^2 \right)\d g_{ij}.
\label{R-gamma}
\ee
Upon using these formulas, we can expand the Wen-Zee action to
quadratic order in the fluctuations of both gravity and
electromagnetic backgrounds, and obtain:
\be
\label{WZmetric}
S_{WZ}= \frac{\n\bar{s}}{4\pi}\int d^3x \left(  \ A_0 \mathcal{R} +  
\ \e^{ij} \dot{A_i} \G_j 
-\frac{B_0}{4}\e^{ij} \d g_{ik}\d \dot{g}_{jk} \right).
\ee

\begin{figure}[h]
\begin{center}
\includegraphics[width=0.6\textwidth]{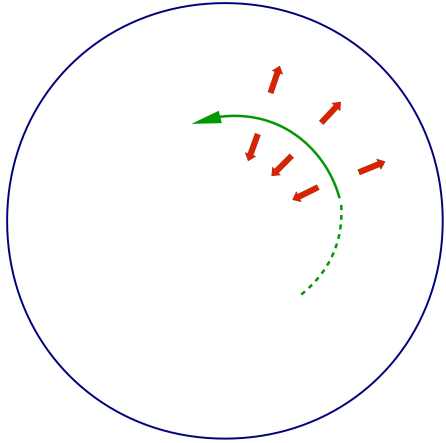}
\caption{Illustration of the Hall viscosity:
a counter-clockwise stirring of the fluid in the bulk of the
droplet causes an orthogonal force (red arrows).}
\label{visco}
\end{center}
\end{figure}

From this expression, we can compute the induced stress tensor to
leading order in the metric and for constant magnetic field 
${\cal B}(x)=B_0$, obtaining the result:
\be
\label{stress-tensor}
T_{ij}=-\frac{2}{\sqrt{g}}\frac{\d S}{\d g^{ij}}= -
\frac{\h_H}{2}\left(\e_{ik}\dot{g}_{kj} + \e_{jk}\dot{g}_{ki} \right),
\ee 
with 
\be 
\label{hallvisco}
\h_H=\frac{\r_0 \bar{s}}{2}=\frac{\n \bar{s}B_0}{4\pi}.
\ee 
The coefficient $\h_H$ is the Hall viscosity: it
parameterizes the response to stirring the
fluid, that corresponds to an orthogonal non-dissipative force 
(see Fig. \ref{visco}) \cite{leigh}.  
Avron, Seiler and Zograf were the first to discuss
the Hall viscosity from the adiabatic response \cite{Avron}, followed
by other authors \cite{Haldane}\cite{Read}\cite{Kletsov}; in
particular, the relation between the Hall viscosity and $\bar s$
(\ref{hallvisco}) has been shown to hold for general Hall fluids
\cite{Read}.

Let us analyze the expression of the stress tensor (\ref{stress-tensor}). 
Being of
first order in time derivatives, it describes a non-covariant effect, in
agreement with the fact that the Wen-Zee action is only covariant
under time-independent coordinate reparameterization and local frame rotations.
At a given time $t=0$ we can choose the conformal gauge for the metric,
$g_{ij}(0,x)=\sqrt{g}\,\d_{ij}$, and consider time-dependent 
coordinate changes, $\d x^i=u^i(t,x)$, representing the deformations. 
These can be decomposed into conformal transformations
and isometries (also called area-preserving diffeomorphisms):
the former maintain the metric diagonal and obey 
$\de_i u_j+\de_j u_i=\d_{ij}\, \de_ku^k $; the latter keep  
its determinant constant and satisfy $\de_k u^k=0$.

The conformal transformations do not contribute to the 
stress tensor (\ref{stress-tensor}); the isometries generated by the 
scalar function $w(t,x)$ yield:
\be
T_{ij}=\h_H\left(2\de_i \de_j- \d_{ij}\de^2 \right) \dot{w},
\qquad\quad 
\d x^i=u^i=\eps^{ij}\,\de_j w(t,x)\ .
\label{T-winf}
\ee
Therefore, we have found that the orthogonal
force is proportional to the shear induced
in the fluid by time-dependent area-preserving diffeomorphisms.

The last effect parameterized by $\bar s$ that we mention in 
this section is a correction to the density and Hall 
current in presence of spatially varying electromagnetic
backgrounds (in flat space). This is given by \cite{Son1}:
\ba
\r&=&\frac{\nu}{2\pi}\left(1- 
\frac{\bar{s}+\bar{s}_o}{2}\frac{\de^2}{B_0} 
+O\left(\frac{\de^4}{B_0^2}\right) \right) {\cal B}(x),
\label{rho-b}
\\
J^i&=&\frac{\nu}{2\pi}\eps^{ij} 
\left(1-\frac{\bar{s}+\bar{s}_o}{2}\frac{\de^2}{B_0}
+O\left(\frac{\de^4}{B_0^2}\right) \right) {\cal E}^j(x).
\label{current-b}
\ea
In these equations, the coefficient $\bar s$ has an additive
non-universal correction $\bar s_o$ that depends on the value of the
gyromagnetic factor in the microscopic Hamiltonian
\cite{Son1}\cite{Abanov1}\cite{Wiegmann1}.  The results (\ref{rho-b},
\ref{current-b}) do not follow from the Wen-Zee action because they
are of higher order in the derivative expansion, i.e.  in the series
$(\de^2/B_0)^n$ involving the dimensionful parameter $B_0$.  They were
obtained by an independent argument in Ref. \cite{Son1}, and later
deduced from the Wen-Zee action upon assuming local Galilean
invariance \cite{Son2}.  Our results in this paper will not rely on
the presence of this symmetry, and we refer to the works
\cite{Son1}\cite{Son2}\cite{Abanov2} for an analysis of its
consequences.

\begin{figure}
\begin{center}
   {\includegraphics[width=0.7\textwidth]{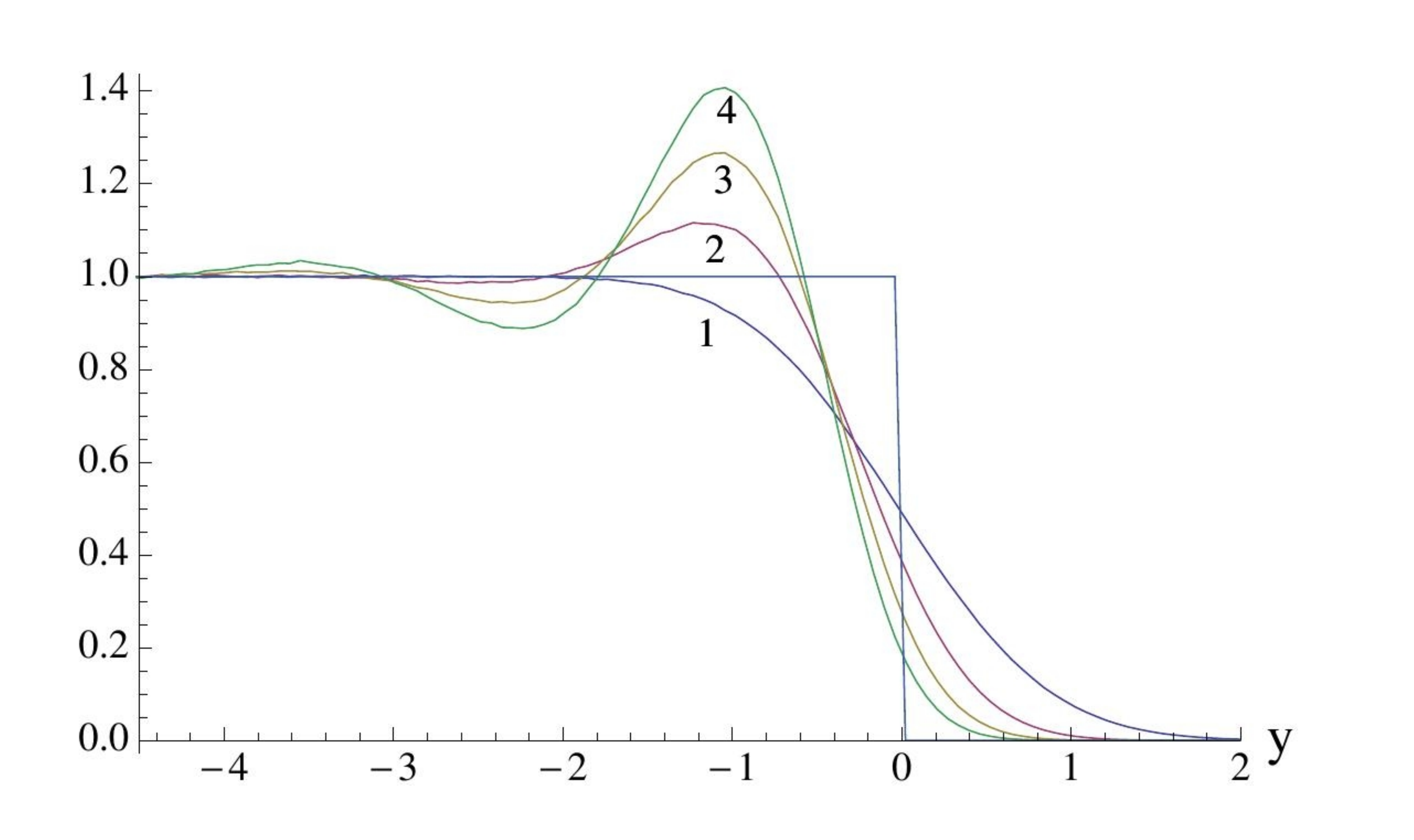}}
   \caption{Numerical density profile of the droplet for the $N=200$ 
electrons Laughlin wavefunction, labeled by the value of $p$ 
for $\n=1/p$, from Ref. \cite{edgenumeric}
(the density is normalized to one in the bulk).} 
\label{bump}
\end{center}
 \end{figure}

The correction (\ref{rho-b}) describes an interesting
property for the density profile of Laughlin fluids.
Numerical and analytical studies of fractional Hall states have found 
a prominent peak, or overshoot, at the edge (see Fig. \ref{bump}) 
\cite{edgenumeric}. 
This is in contrast with the integer Hall case, where the profile
is monotonically decreasing at the edge.
Let us consider the two exact sum rules
obeyed by the density of states in the lowest Landau level,
specializing to the Laughlin case ($\nu=1/p$). They read:
\be
\int d^2x\, \r =N, \qquad\qquad
\int d^2x \frac{x^2}{\ell^2}\, \r = M + N = \frac{pN(N-1)}{2} +N,
\label{sum-rules}
\ee
where $\ell=\sqrt{2\hbar c/eB_0}$ is the magnetic length and $M$ is the total
angular momentum. 
The first sum rule is satisfied by a droplet of constant density 
with sharp boundary, that has the form of a radial step function,
$\r(x)= B_0/2p\pi$ for $x^2<Np\ell^2$, $\r(x)=0$ for $x^2>Np\ell^2$.
However, inserting this droplet form in the second sum rule 
only gives the leading
$O(N^2)$ term. This implies that the sub-leading $O(N)$
contribution depends on the shape of the density at the boundary.

We can repeat the calculation with the improved
expression of $\r(x)$ in (\ref{rho-b}):
we assume that ${\cal B}(x)$ has the profile of the sharp droplet
and compute the sum rules including the $O(\de^2/B_0)$ correction.
Upon integration by parts, this correction vanishes in the
first sum rule, while it correctly yields the sub-leading $O(N)$ contribution
in the second sum rule, 
upon matching the parameters $\bar{s}+\bar{s}_o=p/2 -1$.
Of course, changing the profile ${\cal B}(x)$ from a sharp droplet
can alter this result by an additive constant; this is another
indication that this quantity is not universal.
In conclusion, we have found that the intrinsic angular momentum
parameter $\bar{s}$ also accounts for the fluctuation
of the density profile near the boundary of the droplet. 

%-3----------------------------------------------------------
\section{$W_{\infty}$ symmetry and multipole expansion}

\subsection{Quantum area-preserving diffeomorphisms}

\begin{figure}[h]
\begin{center}
\includegraphics[width=0.7\textwidth]{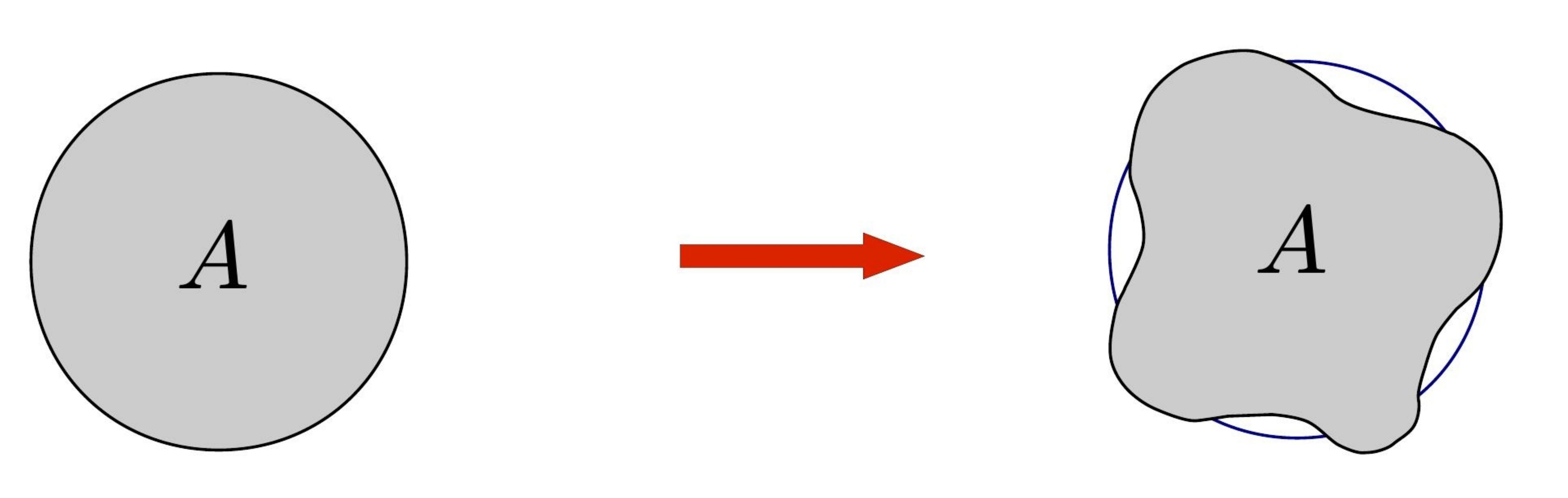}
\caption{\small{Shape deformation of the droplet 
under the action of area preserving diffeomorphisms.}}
\label{areadifffig}
\end{center}
\end{figure}

A droplet of two-dimensional incompressible fluid is characterized at
the classical level by a constant density $\r_0$ and a sharp boundary.
For a circular geometry, the ground state droplet has the shape of a disk 
and fluctuations amount to shape deformations (see Fig. \ref{areadifffig}).
Given that the number of electrons $N=\r_0 A$ is fixed, the area $A$
is a constant of motion, i.e. fluctuations correspond to
droplets of same area and different shapes.  These configurations
of the fluid can be realized by coordinate changes that keep the area
constant, i.e.  by area-preserving diffeomorphisms \cite{W1}.

These transformations, already introduced in (\ref{T-winf}), are
generated by a scalar function $w(t,x)$; the
fluctuations of the density are given by:
\be
\d_w\r=\eps^{ij}\,\de_i\r\, \de_j w=\left\{\r,w\right\},
\qquad\qquad \d x^i=u^i=\eps^{ij}\,\de_j w(t,x)\ ,
\label{area-diff}
\ee
where we introduced the Poisson bracket over the $(x^1,x^2)$ coordinates, 
in analogy with the canonical transformations of a two-dimensional
phase space. The calculation of fluctuations/transformations 
for the ground state density using (\ref{area-diff}) yields
derivatives of the step function that are localized at the
edge, as expected \cite{W1}.

It is convenient to introduce the complex notation for the coordinates,
\be
z=x^1+ix^2,\quad \bar{z}=x^1-ix^2,\quad ds^2=dz\, d\bar z,\qquad
\d_{z\bar z}=\frac{1}{2},\quad \d^{z\bar z}=2,
\label{cc}
\ee
and the corresponding Poisson brackets:
\be
\left\{\r,w\right\}= \eps^{z\bar z}\, \de_z\r\, \de_{\bar z}w +
(z \leftrightarrow \bar z),
\qquad\quad \eps^{z\bar z} =-\eps^{\bar z z}=-2i\ .
\label{pb}
\ee
A basis of generators can be obtained by expanding the function 
$w(z,\bar z)$ in power series,
\be
{\cal L}_{n,m}=z^{n+1}\,\bar  z^{m+1},\qquad\qquad
w(z,\bar z)=\sum_{n,m\ge -1} c_{nm}\, z^{n+1}\,\bar  z^{m+1}\ .
\label{Lnm}
\ee
The ${\cal L}_{n,m}$ generators obey the so-called $w_\infty$ 
algebra of area-preserving diffeomorphisms,
\be
\left\{ {\cal L}_{n,m}, {\cal L}_{k,l} \right\}=
\left((m+1)(k+1)-(n+1)(l+1)\right){\cal L}_{n+k,m+l}\,.
\label{w-algebra}
\ee

We consider now the implementation of this symmetry in the quantum
theory of electrons in the lowest Landau level, where coordinates do
not commute, i.e. $[\hat{\bar{z}},\hat z]=\ell^2$.  
The density and symmetry generators
become one-body operators acting in this Hilbert space, that are
expressed in terms of bilinears of lowest Landau level field operators
$\hat\Psi(z,\bar z)$:
\be
\hat\r=\hat\Psi^\dagger\hat\Psi, \qquad 
\hat{\cal L}_{n,m}=\int d^2z \, \hat\Psi^\dagger(z,\bar z)\, 
z^{n+1}\,\bar z^{m+1}\, \hat\Psi(z,\bar z),
\label{Lnm-ope}
\ee 
Upon using the (non-local) commutation relations 
of field operators, one can find
the quantum algebra of the generators (\ref{Lnm-ope}) \cite{W1},
\ba
\left[ \hat{\mathcal{L}}_{n,m}, \hat{\mathcal{L}}_{k,l}\right]
&=&
\sum_{s=1}^{\text{Min}(m,k)} 
\frac{\ell^{2s}\, (m+1)!(k+1)!}{(m-s+1)!(k-s+1)!s!}\ 
\hat{\mathcal{L}}_{n+k-s+1,m+l-s+1} 
\nl
&& \qquad\qquad - \left(m\leftrightarrow l, \ n \leftrightarrow k \right).
\label{Walgebra}
\ea
This is called the $W_\infty$ algebra of quantum area-preserving 
transformations. The terms on the right hand side form an expansion in 
powers of $\ell^2=2\hbar/B_0$: the first term corresponds to the
quantization of the classical $w_\infty$ algebra (\ref{w-algebra}), while the
others are higher quantum corrections $O(\hbar^n)$, $n>1$.

At the quantum level, the classical density given by the
ground state expectation value 
$\r(z,\bar z) =\bra{\Omega}\hat{\rho}\ket{\Omega}$, becomes a Wigner 
phase-space density function, owing to the non-commutativity of coordinates.
The quantum fluctuations of the density are given by the
commutator with the generator $\hat w$ \cite{Sakita},
\be
\label{Moyal}
\d \r (z,\bar{z})=i \bra{\Omega}[\hat{\rho},\hat{w}]\ket{\Omega}=
i \sum_{n=1}^{\infty}\frac{(2\hbar)^n}{B_0^n \ n!}
\left(\de_{\bar{z}}^n\rho\ \de_{z}^n w-
\de_{\bar{z}}^nw \ \de_{z}^n \r \right) \equiv
\left\{ \r, w \right\}_M,
\ee
where $w(z,\bar z)=\bra{\Omega} \hat{w} \ket{\Omega}$.
The non-local expression on the right-hand side is
called the Moyal brackets $\{\r,w\}_M$. The leading $O(\hbar)$
term is again the quantum analog of the classical transformation 
(\ref{area-diff},\ref{pb}).
These results are well-known in the lowest Landau level physics;
in particular, the algebra of two densities in Fourier
space $\hat\r(k,\bar k)$ is obtained by taking the Moyal brackets
(\ref{Moyal}) of two plane waves, leading to the Girvin-MacDonald-Platzman
sin-algebra \cite{GMP}.
\ba
\label{gmp_a}
&&\bra{\Omega}\left[\hat{\rho}(k,\bar k),
\hat{\rho}(p,\bar p) \right] \ket{\Omega}=
\nl
&&\qquad \qquad = \left\{\r(k,\bar k),\r(p,\bar p) \right\}_M= 2
\sinh\left(\frac{p\bar k-\bar p k}{8} \right)
\r(k+p,\bar k +\bar p)\ .
\ea

The $W_\infty$ symmetry of Laughlin and hierarchical fluids has been
investigated in several works \cite{W1}\cite{W2}, that mainly studied
its implementation in the conformal field theory of edge
excitations. In the limit to the edge, the density and $W_\infty$
generators (\ref{Lnm-ope}) become operators in the $(1+1)$-dimensional
theory of the Weyl fermion $\hat F$. Their expressions are \cite{CDTZ}: 
\be
\hat\r(R\th)=\hat F^\dagger(\th)\hat F (\th), \qquad \hat{\cal
  L}_{n,m}=\oint d\th \, \hat F^\dagger(\th)\, e^{i(n-m)\th}\, \left(i
\frac{\de}{\de\th}\right)^{m+1} \, \hat F(\th),
\label{r-lnm}
\ee 
where $R\th$ is the coordinate on the boundary, with $R$ fixed, 
such that $z \to R\exp(i\th)$ and $z\bar z\sim z\de_z \to i \de_\th$.
Thus, the conformal theory possesses chiral conserved currents
of increasing spin (scale dimension), $\s=0,1,2,\dots$, whose
Fourier components are given by (\ref{r-lnm}). These are: 
the charge $W^0=F^\dag F$, the stress tensor
 $W^1\equiv T=F^\dag\de F$, the spin two field  
$W^2=F^\dag\de^2 F$, and so on \cite{CDTZ}.
The general conformal theories with $W_\infty$ symmetry
include multicomponent fermionic and bosonic theories and certain
coset reductions of them. In particular,
the Jain hierarchy of fractional Hall states was uniquely derived by
assuming this symmetry and the minimality of the spectrum of excitations
\cite{W2}.

%-3.2----------------------------------------------------------

\subsection{Higher spin fields}

The formula (\ref{Moyal}) of the Moyal brackets is the central point of the
following discussion.
It expresses the fact that the fluctuations of the density
are non-local functions of the density itself. This is not
surprising, since any excitation in the lowest Landau level  
cannot be localized in an area smaller than $\pi\ell^2$.
Nevertheless, the non-locality is controlled by the $\hbar$, or $1/B_0$, 
expansion. Let us consider (\ref{Moyal}) to the 
second order in $1/B_0$ ($\hbar=1$):
\ba
\label{expansion1}
\d \r&\sim& \frac{i2}{B_0}\, \de_{\bar{z}} \r \ \de_z w\, +
\frac{2i}{B_0^2}\,  \de^2_{\bar{z}}\r\, \de^2_z w\, +\text{h.c.}
\nl
&=&- \eps^{z \bar z}\left(
\frac{1}{B_0} \de_{\bar z} \left( \r\de_z \tilde w\right) \ +
\  \frac{1}{B_0^2} \de^2_{\bar{z}} \left( \r\de^2_z w \right)
\right) + (z \leftrightarrow \bar z)\ .
\label{w-rho}
\ea
In the second line of this equation, 
we reordered the derivatives and added one scalar term in $w \to \tilde w$.
The tensor structure of this expression involves a spin one field
$(a_z, a_{\bar z})$ and a traceless symmetric tensor field 
$(b_{zz}, b_{\bar z \bar z})$ in two dimensions as follows:
\be
\d \r= \eps^{z \bar z}\,\de_{\bar{z}}\left(
a_z + \frac{1}{2 B_0}\de_{\bar{v}}b_{zv}\d^{\bar v v} + 
\frac{1}{2 B_0}\de_{v}b_{z \bar v}\d^{ v\bar v}  \right) 
+ (z \leftrightarrow \bar z)\ ,
\label{higher1}
\ee
since $b_{v \bar z}=b_{\bar z v}=0$, 
with $v$ another complex variable. The fields $(a_z,b_{zz})$ are independent
because $\r,w$ are general functions; they are also irreducible with respect
to the O(2) symmetry of the plane.

In the first term of \eqref{higher1}, we recognize the zero component 
of the matter current expressed in terms of the hydrodynamic gauge field,
 $j^\mu_{(1)}=\eps^{\mu\nu\r}\de_\nu a_\r$, as discussed in Section two. 
Indeed, the other components $j^i_{(1)}$, involving also $a_0$, are uniquely
determined by the requirements of current conservation and
gauge invariance of $a_\mu$.
The second term in (\ref{higher1}) is similarly rewritten:
\be
j^\mu_{(2)}=\frac{1}{B_0}\eps^{\mu\nu\r}\de_\nu\, \de_k\,  b_{\r k},
\qquad\quad \m,\n,\r=0,1,2, \ \ \ k=1,2,
\label{current2}
\ee
where the components of the spin-two field are 
$b_{\m k}= (b_{01},b_{02},b_{11},b_{12},b_{21},b_{22})$
and the summation over spatial indices $k$ is implicit.
In this expression, the gauge symmetry,
\be
b_{\m k} \to b_{\m k} + \de_{\m}v_k\ ,
\label{gauge2}
\ee
involving the space vector $v_k$, can be used to fix two
space components of $b_{jk}$, making it symmetric and traceless.
Moreover, the two components $b_{0k}$ 
will turn out to be Lagrange multipliers, such that
the field $b_{\m k}$ represents two physical degrees of freedom,
namely the original $(b_{z z},b_{\bar z \bar z})$.

In summary, we can view the expansion (\ref{expansion1}) of the Moyal bracket
as the gauge-fixed time component of the current:
\be
\label{cur_sum}
j^{\m}=j^{\m}_{(1)}+j^{\m}_{(2)}=\e^{\m\n\r}\de_{\n}a_{\r}  \ + 
\frac{1}{B_0}\e^{\m\n\r}\de_{\n}\de_{k}b_{\r k} \ .
\ee
The analysis can be similarly extended to the $O(1/B_0^3)$ term in (\ref{Moyal})
involving the spin three field $c_{\mu kl}$, that is fully symmetric
and traceless with respect to its three space indices, and again possesses
two physical components, $(c_{zzz},c_{\bar z\bar z \bar z})$;
this term will be analyzed in Section 3.5. Continuing the 
expansion one encounters further irreducible higher-spin fields 
that are fully traceless and symmetric.

We conclude that the $W_\infty$ symmetry of the incompressible
fluid in the lowest Landau level shows the existence of non-local 
fluctuations, that can be made local by expanding in powers of $1/B_0$ 
and introducing a generalized hydrodynamic approach with
higher-spin traceless symmetric fields. This is suggestive of
a multipole expansion, where the first term 
reproduces Wen's theory, and the sub-leading terms
give corrections that explore the dipole and higher moments of 
excitations.

We finally remark that in the expression of the Moyal brackets (\ref{Moyal}), 
the coefficients of the quantum terms $O(\hbar^n)$, $n>1$,
may depend on the ground state of the system,
but the general derivative expansion is kept. The $W_{\infty}$ symmetry 
also holds for Hall incompressible fluids that fill a finite number of Landau
levels beyond the first one \cite{W1}.

%-3.3--------------------------------------------------

\subsection{The effective theory to second order}

The construction of the effective theory for the spin-two field $b_{\m k}$
follows the usual steps described at the beginning of Section 2.
We need to couple the current $j^\mu_{(2)}$ in (\ref{current2}) to the external 
field $A_\mu$ and introduce a dynamics for the new field.

The action for $b_{\mu k}$ should possess the gauge symmetry
(\ref{gauge2}), treat the time components $b_{0k}$ non-dynamical and 
possess as much Lorentz symmetry as possible. 
To lowest order in derivatives, the following 
generalized Chern-Simons action satisfies these requirements:
\be
S^{(2)}=-\frac{1}{2\g B_0} \int d^3x \, \e^{\m\n\r} 
\, b_{\m k}\,  \de_{\n} b_{\r k}.
\label{CSb}
\ee
The main difference with the standard action for $a_\mu$ is the
lack of Lorentz symmetry. 

In the search of higher-spin field theories in $(2+1)$
dimensions, we can take advantage of the works \cite{HS}, that have 
introduced the following family of relativistic actions:
\be
S_{CSHS}= \int d^3x \, \e^{\m\n\r} 
\, b_\mu^{\{A_i\}}\,  \de_{\n} b_\r^{\{B_j\}}\, \d_{\{A_i\}\{B_j\}},
\label{CSHS}
\ee
where $b_\m^{\{A_i\}}=b_\mu^{A_1,\cdots, A_{\s -1}}$ is totally symmetric
with respect to its $(\s -1)$ local-Lorentz indices, $A_i=0,1,2$, and
$\d_{\{A_i\}\{B_j\}}$ is the totally symmetric delta function.
The actions (\ref{CSHS}) can be made general covariant and
reduce to $S^{(2)}$ in the non-relativistic limit (for $\s=2$).
In the following, we shall keep the discussion as simple as possible
and derive the effective action to quadratic order in the fluctuations.
In this approximation, we can consider the index $k$ of $b_{\mu k}$ 
as the space part of a local-Lorentz index.
Note also that we do not extend the field $b_{\mu k}\to b_{\mu \nu}$,
totally symmetric in $(\mu\nu)$, because in the action (\ref{CSb}) 
this would imply a canonical momentum for $b_{0k}$  that is not wanted.

The hydrodynamic effective action for $b_k=b_{\mu k}dx^\mu$, including the
electromagnetic coupling $j^\mu_{(2)}A_\mu$  is therefore given by:
\be
S^{(2)}[b,A]=\int -\frac{1}{2\g B_0}b_k\, d\, b_k +
\frac{1}{B_0}
A\, d\, \de_k b_k\ .
\label{Seff2}
\ee
Upon integrating the $b_k$ field, one obtains the following contribution
to the induced effective action (\ref{CS}),
\be
S^{(2)}[A]=-\frac{\g}{2B_0}\int \D A\, d\, A\ ,
\label{Sind2}
\ee
where $\D$ is the Laplacian. 
Therefore, we have obtained the  $O(1/B_0)$ correction to the
density and Hall current for slow-varying fields, 
discussed at the end of Section two, Eqs.(\ref{rho-b}),(\ref{current-b}).

%-3.3.1---------------------------------------------
\subsubsection{Coupling to the spatial metric}

We now introduce a metric background in the limit of
weak gravity and obtain the effective action to quadratic order in
the electromagnetic and metric fluctuations.
We let interact the metric with the $b_{\m k}$ field, independently 
of the $a_\mu$ fluctuations, by defining
the stress tensor $t^{ik}$ that couples to the metric $g_{i k}$, as follows:
\be
\label{hydrostress}
t^{\m k}= \e^{kn}\e^{\m\n\r}\de_{\n} b_{\r n}\ .
\ee
In this expression, we added the component $t^{0k}$ such that the stress
tensor is conserved by construction, $\de_{\m} t^{\m k}=0$.
Regarding the space components, we find that the
anti-symmetric part,
\be
\eps_{ik}t^{ik}=-\eps^{ij}\left(\de_j\, b_{0i}-\de_0\, b_{ji}\right)\ ,
\label{antisym}
\ee
is proportional to the Lagrange multiplier $b_{0i}$ that can be put to zero
on all observables by a gauge choice. Namely, the stress tensor
(\ref{hydrostress}) is symmetric ``on-shell''.

Some insight on the definition of the stress tensor \eqref{hydrostress}
can be obtained by comparing it with the expression 
(\ref{current1}) of the matter 
current $j^\mu_{(1)}$ in terms of the hydrodynamic field $a_\mu$.
The fluctuation of the charge is given by the integration of the density
over the droplet, 
\be
\label{deltaQ}
\d Q =\int_D d^2 x\, \d \r=\oint _{\de_D} dx^i a_i\,.
\ee
This reduces to a boundary integral of the hydrodynamic field,
as expected for incompressible fluids.
Similarly, the integral of the stress tensor gives the momentum 
fluctuation,
\be
\label{deltaP}
\d P^{k}=\int_D d^2 x \ t^{0k}=\e^{kl}\oint_{\de D} dx^i \ b_{il}=u^k,
\ee
that is expressed by the boundary integral of the spin-two 
hydrodynamic field. Further higher-spin fields measure other tensor
quantities at the boundary, thus confirming the picture of the
multipole expansion of the droplet dynamics. 
This argument also gives some indications on the matching 
between higher-spin fields in the bulk and
on the edge (\ref{r-lnm})
(the bulk-edge correspondence will be further discussed in the Conclusions).

%-3.3.2----------------------------------------------
\subsubsection{The Wen-Zee action rederived}

Next, we introduce the metric coupling $\l\d g_{\mu k}\, t^{\mu k}$ in
the second order action (\ref{Seff2}), including an independent
 constant $\l$ and the component $g_{0k}$ for ease of
calculation, to be put to zero at the end:
\be
S^{(2)}[b,A,g]=\int -\frac{1}{2\g B_0}b_k\, d\, b_k +
\frac{1}{B_0} A\, d\, \de_k b_k +
\l\d g_{\mu k}\, \e^{kn}\e^{\m\n\r}\de_{\n} b_{\r n}\ .
\label{Seff2-g}
\ee

After integration of $b_{\m k}$, the induced effective action 
takes the form:
\be
\label{second-mix}
S^{(2)}[A,g]=S^{(2)}_{\text{EM}}[A]+
S^{(2)}_{\text{MIX}}[A,g]+ S^{(2)}_{\text{GR}}[g],
\ee
where the three terms read,
\ba
S^{(2)}_{\text{EM}}[A]&=& 
-\frac{\g}{2 B_0} \int d^3x\ \e^{\m\n\r} \D A_{\m}\de_{\n}A_{\r}\ ,
\label{EM2}
\\
S^{(2)}_{\text{MIX}}[A,g]&=&
-\l\g  \int d^3x \ \e^{ij} \e^{kn} 
\left( A_0 \de_i - A_i \de_0 \right) \de_k \d g_{jn} \ ,
\label{MIX2}
\\
S^{(2)}_{\text{G}}[g] &=&
-  \frac{B_0\g \l^2}{2} \int d^3x \e^{ij} \d g_{ik}\d\dot{g}_{jk}\ .
\label{GR2}
\ea

The first term is the $O(1/B_0)$ electromagnetic correction already found
in (\ref{Sind2}). The second and third terms can be rewritten 
using formulas (\ref{w-gamma}) and
(\ref{R-gamma}) of Section two, as follows:
\be
S^{(2)}_{\text{MIX}}[A,g]+ S^{(2)}_{\text{G}}[g]=
\l\g\int d^3x \left(
A_0 {\cal R}+\eps^{ij}\dot{A}_i \G_j-
\frac{B_0\l}{2}\eps^{ij}\d g_{ik} \d\dot{g}_{jk} \right)\ .
\label{sumWZ}
\ee
We have thus obtained the same expression of the Wen-Zee action 
(\ref{WZmetric}) approximated to quadratic order in the fluctuations.
The parameters are identified as,
\be
\label{constants}
\g=\frac{\n \bar{s}}{2 \pi}, \qquad\qquad \l=\frac{1}{2}\ .
\ee
Equations (\ref{Seff2-g}) and (\ref{sumWZ}) are
the main result of this paper. We have found that the $W_\infty$ symmetry
of incompressible fluids led to introduce
a spin-two hydrodynamic field whose coupling to the metric 
reproduces Wen-Zee result obtained by coupling the spin connection 
to the charge current (cf. Eq. (\ref{grav-coupl})).

%-3.3.3---------------------------------------------

\subsection{Universality and other remarks}

Let us add some comments:
\begin{itemize}
\item The result (\ref{sumWZ}) seems to indicate that the gravitational
interaction through spin (\ref{grav-coupl}) of the Wen-Zee approach is
  equivalent to the coupling to angular momentum of extended
  excitations.

\item Nonetheless, the $W_\infty$ symmetry implies the multipole
  expansion (\ref{Moyal}), whose higher components should yield
  further geometric terms in the effective action (see next Section).

\item In this approach, momentum and charge fluctuations are described by
independent fields, $b_{\mu k}$ and $a_\mu$, respectively.  In the
microscopic electron theory, the fixed mass to charge ratio implies
the relation $P^i=(m/e)J^i$ between the two currents; this fact is at
the basis of the local Galilean symmetry (Newton-Cartan approach) that
has been investigated in the Refs.
\cite{Son1}\cite{Son2}\cite{Abanov2}. However, in the lowest Landau
level $m$ vanishes and the quasiparticle excitations, being composite
fermions or dipoles, could have independent momentum and charge
fluctuations.  In particular, purely neutral excitations at the edge
are present for hierarchical Hall fluids \cite{Wenbook}\cite{W2}.

\item The quadratic action (\ref{GR2}) is invariant under spatial
  time-independent reparameterizations within the quadratic
  approximation. One can easily extend it to be fully space covariant;
  however, we do not understand at present how to consistently treat
  the time-dependent non-covariant effects. In particular, there could
be several extensions, corresponding to a lack 
of universality for the results. This point is left to future
  investigations.

\item The $O(1/B_0)$ correction to the Chern-Simons action
provided by $S^{(2)}_{EM}$ in (\ref{EM2}) is non-universal as already
discussed at the end of Section two. Actually, any addition of terms
involving powers of the Laplacian and of the curvature,
\ba
S[A,g]&=&
\frac{\nu}{4\pi}\int \left[1+\d_1 \frac{\D}{B_0}+\cdots
+\b_1\frac{{\cal R}}{B_0}+\cdots \right]
Ad A 
\nl
&+&
\frac{\nu\bar s}{2\pi}\int \left[1+\d_2 \frac{\D}{B_0}+\cdots
+\b_2\frac{{\cal R}}{B_0}+\cdots \right] Ad\w \, ,
\label{S-laplacian}
\ea
amounts to local deformations that are non-universal (including
also the higher-derivative Maxwell term). They can always
added a-posteriori in the effective action approach and their
coefficients $\d_i,\b_i,\cdots$ can be tuned at will.
In particular, including the Laplacian correction (\ref{S-laplacian})
into the expression (\ref{Sind2}) and comparing with the 
known result (\ref{rho-b}), leads to the parameter matching:
\be
\frac{\nu\bar s}{4\pi}-\frac{\nu\d}{4\pi}=
\frac{\nu(\bar{s}+\bar{s}_0)}{8\pi}\ .
\label{gs-const}
\ee

\item Laplacian and curvature corrections to the density and Hall
  current of  Laughlin fluids have been computed to higher order in
  Refs.\cite{Wiegmann1}. They have been obtained for a clean system
  without distortions and thus should be considered as fine-tuned for
  a realistic setting.

\item In deriving the effective theory for the $b_{\mu k}$ field, we
  have assumed its dynamics to be independent from that of $a_\mu$.
  Actually, a non-diagonal  Chern-Simons term 
$\int b_k\, d\,\de_k a\ $ could be added to the action (\ref{Seff2-g}),
but this would lead to further Laplacian corrections in (\ref{S-laplacian}).
\end{itemize}

%-3.5------------------------------------------------

\subsection{The third-order term}

The third term in the Moyal brackets (\ref{Moyal}),
$\d\r\sim i\de_{\bar z}^3\r\, \de_z^3 w/B^2_0 + {\rm h.c.}$, after reordering
of derivatives let us introduce a spin-three field that is totally
symmetric and traceless in the space indices, with components
$(c_{zzz}, c_{\bar z\bar z\bar z})$: 
\be
\d \r_{(3)}=\frac{1}{B^2_0}\eps^{\bar z z}\de^3_{\bar z}\, c_{zzz} + {\rm h.c.}
\,.
\label{densita3}
\ee
This expression can be considered as the gauge fixed, on-shell expression
of the following current,
\be 
j^{\m}_{(3)}=\frac{1}{B^2_0}\e^{\m\n\r} \de_{\n}\,
\de_k\de_l \, P^{k'l'}_{kl}\, c_{\r k'l'},
\label{current3}
\ee
where
\be
P^{n'l'}_{nl}=\frac{1}{2}\left( \d^{n'}_{n}\d^{l'}_{l}
+\d^{n'}_{l}\d^{l'}_{n} -\d_{nl}\d^{n'l'} \right)\ ,
\label{proj}
\ee
is the symmetric and traceless projector respect to the $(nl)$ 
indices.
In equation (\ref{current3}), the spin-three field $c_{\m kl}$, traceless
symmetric on the $(kl)$ indices, has now six components
$c_{\m kl}=(c_{0zz}, c_{0 \bar z\bar z}, c_{zzz}, c_{\bar z z z},
c_{z\bar z\bar z}, c_{\bar z\bar z\bar z})$. Two of them can be
fixed by the gauge symmetry,  $c_{\m kl} \to c_{\m kl} + \de_{\m}v_{kl}$,
with traceless symmetric $v_{kl}$, while the two components with time index are
Lagrange multipliers, leading again to two physical components.

The natural form of the coupling of the spin-three field 
to the metric, although not uniquely justified, is the 
same as that of the spin-two field
(\ref{hydrostress}) with an additional derivative:
\be
t^{\m k}_{(2)}=\frac{1}{B_0}\e^{kn}\e^{\m\n\r}\de_{\n}\de_l \ 
P^{n'l'}_{nl}c_{\r  n'l'},
\label{stress2}
\ee
The kinetic term for the spin-three field with the desired gauge symmetry
and other properties has again the generalized Chern-Simons form (\ref{CSHS}).
In summary, the third-order effective hydrodynamic action 
is ($c_{kl}=c_{\mu kl}dx^\mu)$:
\be
\label{Seff3}
S_{(3)}[c,A,g]=\int -\frac{1}{2\a B^2_0}
c_{kl}\, d\, c_{kl} + A_\mu j^\mu_{(3)} + \eta\, g_{\mu k} t_{(2)}^{\mu k}\ .
\ee

The integration over the spin-three field yields the following
induced effective action,
\be
\label{third-mix}
S^{(3)}[A,g]=S^{(3)}_{\text{EM}}[A]+
S^{(3)}_{\text{MIX}}[A,g]+ S^{(3)}_{\text{GR}}[g],
\ee
where:
\ba
S^{(3)}_{EM}[A]&=& \frac{\a}{4B^2_0}\int \D^2\, AdA \,,
\label{EM3}\\
S^{(3)}_{MIX}[A,g] &=& -\frac{\a\eta}{2B_0}\int
 A_0 \D {\cal R}  +\eps^{ij} \dot{A}_i \D \G_j \,,
\label{MIX3}\\
S^{(3)}_{GR}[g] &=& \frac{\a\eta^2}{4}\int 
\eps^{ij} \d g_{ik} \D \d\dot{g}_{jk}\,.
\label{GR3}
\ea

We thus obtain local Laplacian corrections to the same terms that
occur in the second-order action (\ref{EM2})-(\ref{GR2}).
This is not surprising because both couplings in (\ref{Seff3})
are derivatives of the lower-order ones (\ref{Seff2-g}).

It is natural to compare the result (\ref{GR3}) with the gravitational Wen-Zee
action in (\ref{CS-WZ})
\ba
S_{GRWZ}[g] &=&
\xi\int \w\, d\, \w =\xi\int \left(\w_0 {\cal R}- \eps^{ij}\w_i\, 
\dot{\w}_j \right)
\label{GRWZ2}\\
&\sim& \frac{\xi}{4}\int \eps^{ij}\d g_{ik} 
\left( \d_{kl} \D-\de_k\de_l \right) \d \dot{g}_{jl} \,,
\label{GRWZ1}
\ea
where $\xi=\left(\nu \bar{s}^2- c/12 \right)/4\pi$. In the second
line of this equation we also wrote the expansion to quadratic
order in the fluctuations, to which the cubic term $\w_0 {\cal R}$
does not contribute.

Equation \eqref{GRWZ1} shows that the gravitational Wen-Zee term contains
Laplacian and curvature corrections
to the Hall viscosity (\ref{stress-tensor}).
The comparison with the $W_\infty$ result (\ref{GR3}) shows that 
the expressions of $S^{(3)}_{GR}$  and $S_{GRWZ}$ are similar but
not identical, to quadratic order. The explicit calculation
of the induced action for integer filling fractions of Ref.\cite{Abanov1}
is in agreement with (\ref{GR3}). Following the discussion of universality
in Section 3.4, we are lead to conclude that the Laplacian
corrections in the third-order 
$W_\infty$ action (\ref{EM3}-\ref{GR3}) and the gravitational Wen-Zee
term (\ref{GRWZ1}) are non-universal.
We further remark that the curvature correction $\int \w_0 {\cal R}$ in
(\ref{GRWZ2}), not obtained in our approach, is believed to be universal
because it is also found in the calculation of the Hall viscosity from
the Berry phase of the Laughlin
wavefunction in a curved background \cite{Kletsov}.

 \begin{figure}
 \centering
   {\includegraphics[width=0.6\textwidth]{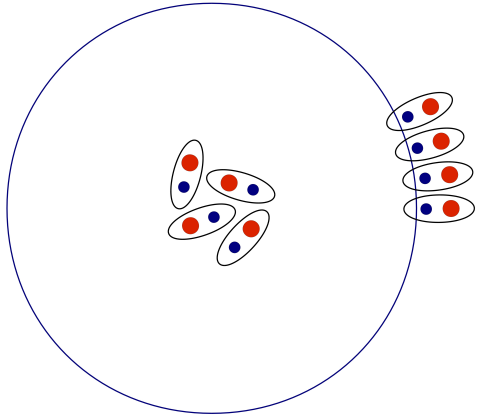}}
   \caption{Dipoles aligned at the boundary.}
   \label{over1}
 \end{figure}

%-4---------------------------------------
\section{The dipole picture}

We now present some heuristic arguments that explain two results of
the previous sections in terms of simple features of dipoles.

The first observation concerns the fluctuation of the density profile at the
boundary (Fig.\ref{bump}). We assume that the low-energy excitations
of the fluids are extended objects with a dipole moment; their charge is 
not vanishing but takes a fractional value due to the unbalance of the two 
charges in the dipole (numerical evidences of dipoles were first discussed 
in Ref.\cite{JK}, to our knowledge). 
The dipole orientations are randomly distributed in
the bulk of the fluid such that they can be approximated by point-like 
objects with fractional charge (see Fig. \ref{over1}). However, near the 
boundary of the droplet, there is a gradient of charge between the interior 
and the empty exterior; thus, the dipoles align their positive charge
tip towards the interior and create the ring-shaped density 
fluctuation that is observed at the boundary.
The effect is stronger for higher dipole moment, that is
proportional to $\bar s =p/2$, as seen in Fig. \ref{bump}.

\begin{figure}
 \centering
   {\includegraphics[width=0.6\textwidth]{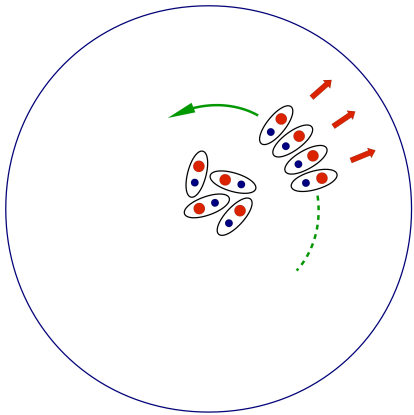}}
   \caption{Hall viscosity caused by dipoles aligned along the 
fluid stream.}
   \label{over2}
 \end{figure}

The second effect that can be interpreted in terms of dipoles is
the Hall viscosity itself (see Fig. \ref{over2}). Again the randomly
oriented dipoles in the bulk are perturbed by stirring the fluid,
namely they acquire an ordered configuration due to the mechanical
forces applied. Any kind of ordered configuration of dipoles,
such as that depicted in the figure, creates a ring-shaped fluctuation
of the density and thus an electrostatic force orthogonal to the 
fluid motion. This effect is parameterized by the Hall viscosity
as discussed in Section two (cf. Fig. \ref{visco}).

The dipole configurations can be matched with the
higher-spin field expansion of the density in 
(\ref{cur_sum}), (\ref{current3}). We rewrite it as follows:
\be
\d \r=\eps^{ij}\de_i\left(
a_j+\frac{1}{B_0}\de_k \, b_{jk} +
\frac{1}{B_0^2}\left(\de_k\de_l -\frac{1}{2}\d_{kl}\D\right)c_{jkl}
+\cdots \right),
\label{dipo-exp}
\ee
and compare to explicit charge configurations.
First consider a bulk charge excitation, $\d \r(\vec x)=q\, \d^2(\vec x)$:
this is parameterized by the leading hydrodynamic field
$a_i\sim O(1/|\vec x|)$, as also shown by (\ref{deltaQ}).
The higher-spin fields do not contribute because they decay faster at
$|x|$ infinity, respectively $\de_k b_{jk}\sim O(1/|\vec x|^2)$ and 
 $\de_k \de_l c_{jkl}\sim O(1/\vec x|^3)$, due to the higher derivatives.
Next, we analyze a dipole configuration,
\be 
\d \r(\vec x)=q \left(\d^2(\vec x+\vec d)-\d^2(\vec x-\vec d)\right),
\label{dipo-ex}
\ee
that corresponds to a $O(1/|\vec x|^2)$ field, for $|\vec x|\gg |\vec d|$.
In this case, both
$a_j$ and $b_{jk}$ contribute. It follows that higher moments of 
the charge configuration gradually involve fields of higher spin values.

We remark that this many-to-one field expansion is a solution for
the non-locality of the dynamics. One could consider a redefinition
of the expansion (\ref{dipo-exp}) in terms of a single field, such as
$u_i=a_i+(1/B_0)\de_k b_{jk}+\cdots$, but this would imply non-local
terms in the Chern-Simon actions (\ref{Seff2-g}), (\ref{Seff3}).
Actually, a non-local formulation of Hall physics based on non-commutative
Chern-Simons theory has been proposed in Ref. \cite{Susskind}, that
corresponds to matrix quantum mechanics and matrix
quantum fields \cite{matrix}.
We think that the present higher-spin approach shares some features 
with the non-commutative theory, while being more general and flexible.

%-5--------------------------------------
\section{Conclusions}

In this paper, we have used the $W_\infty$ symmetry of
quantum Hall incompressible fluids to set up a power expansion
in the parameter $\hbar/B_0$. This analysis leads to a generalized
hydrodynamic approach with higher-spin gauge fields, that
can be interpreted as a multipole expansion of the extended
low-energy excitations of the fluid. To second order, the spin-two
field with Chern-Simons dynamics and 
electromagnetic and metric couplings reproduces the Wen-Zee action.
The third-order term yields non-universal corrections to it.

Regarding the universality of terms of the effective
action, we have pointed out that local gradient and curvature corrections
are non-universal. 
The universal terms and coefficients can be identified with those
that have a correspondence with the conformal field theory on the edge
of the droplet. As is well known, the Chern-Simons terms in the 
effective action,
\be
\label{Wexpansion}
S[a,b,c,\cdots; A,g,\cdots ]=-\int \frac{\pi}{\n} a\, d\,a + 
\frac{\pi}{\nu\bar s B_0} b_k\, d\, b_k + 
\frac{1}{2\a B_0^2} c_{kl}\,d\,c_{kl} +
 \cdots + \text{couplings},
\ee
are not fully gauge invariant and boundary actions are needed
to compensate \cite{Wenbook}. 

Typically, the bulk fields define boundary fields that express the 
boundary action and have spin reduced by one: as is well known,
the field $a_\mu$ defines through the relation $a_\mu =\de_\mu \varphi$
the scalar edge field $\varphi$ that expresses the chiral Luttinger liquid
action \cite{Wenbook}. Namely, the boundary field is the gauge degrees of 
freedom that becomes physical at the edge.
Similarly, the spin-two field identifies an edge chiral vector,
$b_{\mu \th}=\de_\mu v_\th$, with $\th$ the azimuthal direction; 
the spin-three a two-tensor and so on.
It follows that the couplings $\n, \bar s, \a,\cdots$ in 
(\ref{Wexpansion}) also appear as parameters in the edge action and can be
put in relation with observables of the conformal field theory.
Since their values can be related to universal quantities at the edge,
these parameters can be defined globally on the system and manifestly do
not depend on disorder and other local effects. A hint of this
correspondence is already apparent
in the quantities (\ref{deltaQ}), (\ref{deltaP}) discussed in Section 3.3.1.
Let us also mention the work \cite{Abanov-boundary} studying the
boundary terms of the Wen-Zee action.

The analysis presented in this paper could be developed in many aspects:
\begin{itemize}
\item The bulk-edge correspondence for higher-spin actions (\ref{Wexpansion})
should be developed in detail, and the observables of the conformal 
field theory should be identified that express the universal
parameters.
Clearly, the higher-spin fields do not have an independent dynamics at
the edge: for Laughlin states, the higher-spin
currents are expressed as polynomials of the charge current $\de_\th\varphi$
\cite{W-dyna}.

\item The third order effective action could encode universal effects 
if the spin-three hydrodynamic field is coupled to a novel spin-three 
background `metric', the two fields being related by a Legendre 
transform. At present we lack a geometric understanding of
this and higher-spin background fields, and the physical effects that
they describe.

\item The analysis presented in this work should be put in
  contact with the Haldane approach of parametric variations of the Laughlin
  wavefunction, that also involves a traceless spin-two field
\cite{Haldane}. Further deformations could be encoded in the
higher-spin background fields mentioned before. Moreover, 
  our approach should be related to the Wiegmann generalized 
  hydrodynamics of electron-vortex composites \cite{Wiegmann-vortex}.

\item The higher-spin Chern-Simons theories (\ref{Wexpansion}) predict
  new statistical phases for dipole monodromies that require physical
  understanding and verification in model wavefunctions.

\item The whole analysis can be extended to the hierarchical 
Hall states that are described by multicomponent hydrodynamic
Chern-Simon fields
\cite{Wenbook}.

\end{itemize}

{\bf Acknowledgments}

The authors would like to thank A. G. Abanov, A. Gromov,
F. D. M. Haldane, T. H. Hansson, K. Jensen, D. Karabali, S. Klevtsov, 
V. P. Nair, D. Seminara, D. T. Son,
P. Wiegmann and G. R. Zemba for very useful scientific exchanges.
A. C. acknowledge the hospitality and support by the Simons Center for
Geometry and Physics, Stony Brook, and the G. Galilei Institute for
Theoretical Physics, Arcetri, where part of this work was done. The
support of the European IRSES grant, ‘Quantum Integrability, Conformal
Field Theory and Topological Quantum Computation’ (QICFT) is also
acknowledged.

%-A---------------------------------------
\appendix
 \section{Curved space formulas}
 \label{WZflat}
We consider a spatial metric $g_{ij}=g_{ij}(x^k,t)$, with $i,j,k=1,2$, 
depending on space and time and assume that $g_{00}=g_{ij}=0$.
This metric can be written in terms of the spatial zweibeins $e^a_i$ as 
follows,
\be
\label{g-e}
g_{ij}=e^a_i e^b_j\, \d_{ab},
\ee
with the coordinates and local frame indices 
taking the values $i,j,a,b=1,2$. The zweibeins $e^a_i$ and their inverses 
$E^i_a$ satisfy the conditions:
\be
\label{inverse}
 E^i_a e^a_j=\d^i_j, \quad\qquad E^i_a e^b_i=\d^a_b.
\ee
We also assume that the matrix of vielbeins $e^A_\mu$ 
in three dimensions ($\mu,A=0,1,2$),  has
vanishing space-time and time-time components.

When the gravity background has vanishing torsion, the spin 
connection can be expressed in terms of the vielbeins 
\cite{GRbook}. Starting from the three-dimensional expression
($\m,\n,\s=0,1,2$ and $A,B,C=0,1,2$),
 \be
 \label{levi-civita}
  \w_{\m}^{AB}(e)=\frac{1}{2} \left(  E^{\n[ A} \de_{[\m} e^{B]}_{\n]} - 
E^{\n [A} E^{B] \s} e_{C \m} \de_{\n} e^{C}_{\ \s}         \right).
 \ee
and the definition,
 \be
 \label{ominv}
\w^{C}_{\m}=\frac{1}{2}\e^{ABC}\w_{\m AB},
 \ee
we obtain the following results:
\be
\w^a_{\m}=0, \ \  \ \ a=1,2,
\label{w0}
\ee
\be
\label{omega-0}
\w_0 \equiv\w^0_0=\frac{1}{2} \e^{ab}E^{a j} \de_0 e^b_j,
\ee
and
\be
\label{omega-i}
\w_i\equiv\w^0_i=\frac{1}{2} \e^{ab}E^{a j} \de_i e^b_j -
\frac{1}{2}\frac{\e^{jk}}{\sqrt{g}}\de_jg_{ki},
\ee
where $g=\text{det}(g_{ij})$. In the last equation, $\e^{jk}$ 
is the antisymmetric symbol of coordinate space, $\e^{12}=1$, that is 
related to that in local frame space as follows:
\be
\label{eps-eps}
\e^{ab}=\frac{e^a_i e^b_j \e^{ij}}{\sqrt{g}}, \ \ \ \ \ \ \  \ 
\e^{ab}E^{ai}E^{bj}=\frac{1}{\sqrt{g}}\e^{ij}.
\ee 
In two spatial dimensions the Riemann 
tensor $R^{ab}_{ij}$ and the Ricci scalar $\mathcal{R}$ depend on 
the spin connection through the formulas,
\be
\label{R-om}
R^{ab}_{ij}=\left( \de_i \w_j-\de_j\w_i\right)\e^{ab}, \ \ \ \ \ \ 
\mathcal{R}=2\frac{ \de_i \w_j \e^{ij}}{\sqrt{g}}\ .
\ee
Their coordinate components are written in terms 
of the Christoffel symbols $\G^i_{jk}$ as follows:
\be
\label{R-gam}
R^{i}_{jkl}=\de_{j}\G^{i}_{kl} + 
\G^{i}_{jr}\G^{r}_{kl} - (j \leftrightarrow k), \ \ \ \ \ \ \ \ 
\mathcal{R}=g^{jl}R^{k}_{jkl},
\ee
where
\be
\label{gamma}
\G^{i}_{jk}=\frac{1}{2}g^{il}\left(\de_{j}g_{lk}+ \de_{k}g_{lj} - 
\de_{l}g_{jk} \right).
\ee
Finally, in curved space the expression for the magnetic field becomes:
\be
{\cal B}=\frac{\e^{ij}\de_iA_j}{\sqrt{g}}.
\label{Bdef}
\ee

We now find the approximate formulas for small fluctuations around
the flat metric, i.e. $g_{ij}=\d_{ij} + \d g_{ij}$.
Then, $\sqrt{g}\simeq 1$ and $\d g^{ij}=-\d g_{ij}$. 
Choosing a gauge for the local
O(2) symmetry such that the zweibeins form a symmetric matrix, we
find from \eqref{g-e} that:
\be
\label{e-linear}
\d g_{ij}=\d e^a_j \d_{ai} +\d e^a_i \d_{aj} =2 \d e_{ij}.
\ee
In this limit, an approximate expression for $\w_0$ in (\ref{omega-0})
is obtained by making use of \eqref{inverse} and \eqref{e-linear}:
\be
\label{omega-0g}
\w_0=-\frac{1}{8} \e^{ik} \d g_{ij} \d \dot{g}_{kj}.
\ee 
To the linear order, we also find that $\w_j$ in \eqref{omega-i}, 
$\G^{i}_{jk}$ in \eqref{gamma}  
and the Ricci  scalar $\mathcal{R}$ in \eqref{R-om} and 
\eqref{R-gam} take the following expressions:
\be
\label{omega-ig}
\w_j=\frac{1}{2}\e^{ki}\de_i \d g_{kj},
\ee
\be
\label{g-linear}
\G^{i}_{jk}\sim \G_{i,jk}=\frac{1}{2}\left(\de_{j}\d g_{ik}+ \de_{k}\d g_{ij} - 
\de_{i}\d g_{jk} \right),
\ee
\be
\label{r-linear}
\mathcal{R}=
\left(\de_i\de_j - \d_{ij}\de^2 \right)\d g_{ij}.
\ee

%-B-------------------------------------------- 

\end{document}